\documentclass[fleqn,usenatbib]{mnras}

\usepackage{newtxtext,newtxmath}

\usepackage[T1]{fontenc}
\usepackage{ae,aecompl}

\usepackage{graphicx}	
\usepackage{amsmath}	
\usepackage{amssymb}	

\usepackage{color,units}

\title[]{Dark matter model favoured by reionization data: \\ 7~keV sterile neutrino vs cold dark matter}

\author[A.~Rudakovskyi et al.]{A.~Rudakovskyi,$^{1,2,3}$
\thanks{E-mail: rudakovskyi@bitp.kiev.ua} and
D.~Iakubovskyi$^{4,1}$
\\
$^{1}$Bogolyubov Institute of Theoretical Physics, Metrologichna Str. 14-b, 03143, Kyiv, Ukraine\\
$^{2}$Taras Shevchenko National University of Kyiv, Physics Department, Glushkova ave. 2, Kyiv, Ukraine\\
$^{3}$Main Astronomical Observatory, 
Akademika Zabolotnoho str.\@ 27, 03143, Kyiv, Ukraine\\
$^{4}$Discovery Center, Niels Bohr Institute, Blegdamsvej 17, Copenhagen, Denmark\\}

\date{Accepted XXX. Received YYY; in original form ZZZ}

\pubyear{2018}

\begin{document}
\label{firstpage}
\pagerange{\pageref{firstpage}--\pageref{lastpage}}
\maketitle

\begin{abstract}
{
One of possible explanations of a faint narrow emission line at 3.5~keV reported in our Galaxy, Andromeda galaxy and a number of galaxy clusters is the dark matter made of 7~keV sterile neutrinos. Another signature of such sterile neutrino dark matter could be fewer ionizing sources in the early Universe (compared to the standard `cold dark matter' (CDM) scenario), which should affect the reionization of the Universe. By using a semi-analytical model of reionization, we compare the model predictions for CDM and two different models of 7~keV sterile neutrino dark matter (consistent with the 3.5~keV line interpretation as decaying dark matter line)  with available observations of epoch of reionization (including the final measurements of electron scattering optical depth made by \emph{Planck} observatory). We found that both CDM and 7~keV sterile neutrino dark matter well describe the data. The overall fit quality for sterile neutrino dark matter is slightly (with $\Delta \chi^2 \simeq 2-3$) better than for CDM, although it is not possible to make a robust distinction between these models on the basis of the given observations.}
\end{abstract}

\section{Introduction}

To this end, the constituents of dark matter --- the largest gravitating substance in the Universe --- have not been identified. A possible clue on the dark matter origin is the faint narrow emission line-like feature at 3.5~keV reported in our Galaxy, M31 and galaxy clusters. At the moment, there is an ongoing debate about the line status, but according to recent reviews~\citep{Adhikari:16,Abazajian:17, Boyarsky:18}, it can be interpreted as a signal from decaying dark matter, e.g., in the form of right-handed (`sterile') neutrinos with 7~keV mass and the mixing angle with Standard Model neutrinos $\mathrm{sin}^2 2\theta = (2-20)\times 10^{-11}$. 

For such parameters, sterile neutrino dark matter would originate in the early Universe from resonant oscillations of usual left-handed (`active') neutrinos~\citep{Shi:98,Abazajian:01a,Laine:08,Abazajian:14,Venumadhav:15}. As a result, sterile neutrino dark matter, unlike cold dark matter (CDM), would be initially ultra-relativistic with non-thermal distribution function, smearing out density perturbations at small spatial scales. Such smearing, indeed, can be a result of warm dark matter with a few keV mass~\citep[see e.g.][]{Dolgov:00,Abazajian:01b} or of a mixture of cold and warm dark matter~\citep{Boyarsky:08c,Maccio:12b}. It can be traced by a number of observations related to structure formation at different redshifts, such as Lyman-alpha forest power spectrum~\citep{Narayanan:00,Hansen:01,Viel:05,Viel:06,Viel:07,Viel:13,Abazajian:05,Seljak:06,Boyarsky:08c,Boyarsky:08d,Garzilli:15,Irsic:17,Yeche:17,Baur:17}, reionization of the Universe~\citep{Barkana:01,Yoshida:03,Somerville:03,Jedamzik:06,Yue:12,Schultz:14,Dayal:15,Rudakovskyi:16,Bose:16,Cen:16,Lopez-Honorez:17}, subhalo counts in the Local Group~\citep{Maccio:09,Polisensky:10,Lovell:13,Kennedy:13,Horiuchi:13,Lovell:15,Lovell:16a,Lovell:16b,Cherry:17}, 
luminosity functions at low~\citep{Menci:16,Menci:17b} and high~\citep{Song:09,Schultz:14,Corasaniti:16,Menci:17a} redshifts, substructure counts in gravitational lensing systems~\citep{Zentner:03,Miranda:07,Inoue:14,Birrer:17}, galaxy velocity function~\citep{Klypin:14,Schneider:16}, stellar mass -- halo mass relation of isolated field dwarf galaxies~\citep{Read:16}, stellar mass functions at redshifts $z \lesssim 3.5$ together with the Tully--Fisher relation~\citep{Kang:12}, star-formation history of the Local Group dSphs~\citep{Chau:16}, and number density of direct collapse black hole hosts~\citep{Dayal:17}.

In this paper, we study the difference between the 7~keV sterile neutrino dark matter (that can be responsible for the origin of 3.5~keV emission line) models and the standard cold dark matter model on the reionization of the Universe.
Some of previous works~\citep{Rudakovskyi:16,Lopez-Honorez:17} showed that the observational data on reionization may be described better in sterile neutrino dark matter or thermal-relic warm dark matter models compared to the CDM model. The goal of the present paper is to \emph{quantify} this difference using available observations. This paper is organized as follows: Sec.~\ref{sec:method} contains a description of our method and the observations (including the final measurements of electron scattering optical depth made by \emph{Planck}); the obtained results are summarized in Sec.~\ref{sec:results} and discussed in Sec.~\ref{sec:discussion}. Finally, our conclusions are summarized in Sec.~\ref{sec:conclusion}.

\section{Method}\label{sec:method}

To calculate the ionized volume-filling fraction $Q_\text{II}(z)$ and the CMB electron scattering optical depth $\tau_\text{es}$, we used the extension of the `bubble model'~\citep{Furlanetto:04, Yue:12} of reionization, see~\cite{Rudakovskyi:16} for more detailed description. 
We assume that the main source of ionizing photons are
stars formed in galaxy-size dark matter haloes, while smaller haloes (starting from the Jeans mass) work as `recombination sinks' of ionizing radiation due to their higher hydrogen density. At each redshift, we calculate the fraction of haloes that contain ionization or recombination sources by using extended Press-Schechter formalism~\citep{Press:74, Bond:91, Bower:91, Lacey:93}. By solving numerically the main equation of the `bubble model' which relates the mass of ionized gas, the mass of recombined hydrogen and the mass of baryons collapsed into galaxies, we calculated the threshold $\delta_x$ for initial mass overdensities as a function of redshift $z$ and the variance $\sigma^2$ of power density spectrum. Then, approximating threshold as a linear function of $\sigma^2$, $\delta_x(z, \sigma^2) \approx B_0 + B_1\sigma^2$, we obtain an analytic expression for the halo mass function, finally used to calculate the ionizing volume filling fraction $Q_\text{II}(z)$.

The model contains several input parameters:
\begin{itemize}
\item linear dark matter power spectrum $P(k)$, e.g. for CDM or sterile neutrino dark matter;
\item minimum `virial temperature' $T_\text{vir}$  of dark matter halos that host stars responsible for reionization, see~\cite{Haiman:00b, Barkana:00} and references therein;
\item ionizing efficiency $\zeta$, which is the number of ionizing photons released by stars in galaxies per baryon collapsed into DM halos;
\item recombination efficiency $\xi$, which is the average number of recombinations per atom in collapsed
minihaloes \citep{Haiman:00,Iliev:04} during the whole epoch of reionization.
\end{itemize}
We focused on three dark matter models: cold dark matter (CDM) and two models of sterile neutrino dark matter able to explain the observed properties of the 3.5~keV line --- the model L12 (sterile neutrino generated with lepton asymmetry $L_6=12$, that corresponds to sterile neutrino mixing angle $\mathrm{sin}^2 2\theta \simeq 1.6\times 10^{-11}$, see Fig.~1 of~\cite{Lovell:15}) and the model s228899 (sterile neutrino with mixing angle $\mathrm{sin}^2 2\theta=2.8899 \times 10^{-11}$, see~\cite{Horiuchi:15}). Our choice of sterile neutrino dark matter parameters is in agreement with the recent structure formation and X-ray constraints \citep{Baur:17, Boyarsky:18}. 

 The value of $T_\text{vir}$ is related to the minimal mass $M_{vir}$ of dark matter halos which host stars~\citep{Barkana:00}:
\begin{multline}M_{vir} = 1.0\times 10^8 \left(\frac{1+z}{10}\right)^{-3/2}\left(\frac{\mu}{0.6}\right)^{-3/2}\left(\frac{T_\text{vir}}{1.98\times 10^4~\unit{K}}\right)^{3/2}\\\times\left(\frac{\Omega_m}{\Omega_m^z}\frac{\Delta_c}{18\pi^2}\right)^{-1/2}\unit{M_\odot/h},
\end{multline}
where $z$ is the halo redshift, $\mu \simeq 0.60$ is the mean molecular weight, $\Omega_m^z = 1-\Omega_\Lambda/[\Omega_m(1+z)^3+\Omega_\Lambda]$ and $\Delta_c = 18\pi^2 + 82(\Omega_m^z - 1) - 39(\Omega_m^z - 1)^2$~\citep{Bryan:97}.  
According to Sec.~3.3 of~\cite{Barkana:00}, hydrogen cooling becomes efficient for $T_\text{vir} \gtrsim 10^4$~K. In this paper, we fixed  $T_\text{vir} = 10^4$~K  similarly to~\cite{Barkana:00, Furlanetto:04,Yue:12,Rudakovskyi:16}.
Finally, we assume  the ionizing efficiency and the recombination efficiency to vary within the very wide ranges $\zeta = 5$$-$$100$ (which is in agreement with \cite{Greig:16a})\footnote{Note that our definition of $\zeta$ is similar to \cite{Rudakovskyi:16} and doesn't include average number of recombinations in the Universe, unlike, e.g., \cite{Furlanetto:04, Lopez-Honorez:17}.} and $\xi = 0$$-$$9$ (according to \cite{Iliev:04})\footnote{Our definition of $\xi$ is related to definition by \cite{Iliev:04} as $\xi=\xi_\emph{Iliev}-1$.}).%

For an up-to-date summary of observational constraints on reionization history, see, e.g., \cite{Mitra:15,Bouwens:15,Robertson:15,Greig:16a,Konno:17}. However, many of the measurements reported in these papers were obtained by assuming some particular model of reionization. Since our goal is to compare reionization in \emph{different} dark matter models --- cold dark matter (CDM) and 7~keV sterile neutrino dark matter potentially responsible for the narrow line at 3.5~keV, --- it is important to use  measurements that are fully or almost fully model-independent, and to quantify the maximal level of uncertainty for the second case. Therefore, we constructed the following extension of the `Gold Sample' of~\cite{Greig:16a} (all error bars are quoted at 1$\sigma$ level) further used in our modeling:
\begin{itemize}
\item the final value of the electron scattering optical depth $\tau_\text{es} = 0.054 \pm 0.007$ obtained from the combination of the \textit{Planck} temperature correlations (TT) and $E$-mode polarization (EE) correlations at low multipoles $l = 4$$-$$20$), see~\cite{Planck-2018}. According to~\cite{Mesinger:12}, this value is only approximately model-independent in case of `patchy' reionization: since regions with higher electron density become reionized earlier, the all-sky averaged value of $\tau_\text{es}$ can increase by $\lesssim 4\%$, or by $\lesssim 0.2\sigma$, so we assume the value of $\tau_\text{es}$ to be the same for
sterile neutrino models and CDM.

\item the lower bounds of  the ionized 
volume-filling fraction (together with their lower $1\sigma$ errorbars) $Q_\text{II} \geq 0.96 - 0.05$, $Q_\text{II} \geq 0.94 - 0.05$ and $Q_\text{II} \geq 0.62 - 0.20$  obtained  from the model-independent analysis of `dark pixel' fraction in QSO spectra~\citep{McGreer:14} at redshifts 5.48--5.68, 5.77--5.97 and 5.97--6.17, respectively.
\item the value of $Q_\text{II}$ obtained from the analysis of Lyman-$\alpha$ damping wing in the spectra of the quasars ULASJ1120+0641~\citep{Greig:16b} and ULASJ1342+0928~\citep{Greig:18}. To convert the Lyman-$\alpha$ damping wing observations into $Q_\text{II}$, \citep{Greig:16b,Greig:18} used  different models where the driving sources for reionization are faint galaxies (with halo masses $10^8$$-$$10^9~M_\odot$) and bright galaxies (with halo masses $\sim 10^{10}~M_\odot$). For these models, one obtained $Q_\text{II} = 0.60^{+0.19}_{-0.21}$ and $0.54^{+0.21}_{-0.21}$ at redshift $z=7.1$, and $Q_\text{II} =0.79^{+0.19}_{-0.17}$ and $0.72^{+0.23}_{-0.20}$ at redshift $z=7.5$. The observed difference between these two models is about 0.3$\sigma$, and the expected difference between the CDM and sterile neutrino models is even less~\citep[see Discussion in][]{Rudakovskyi:16}, so we constructed the average value $Q_\text{II} = 0.56 \pm 0.23$ for $z=7.1$, and $Q_\text{II}=0.75\pm 0.23$ for $z=7.5$ (by incorporating the error bars from both models) to be the same for CDM and sterile neutrino dark matter.
\end{itemize}

For each dark matter model of our interest (CDM, L12, s228899), we calculate the best-fit values of $\zeta$ and $\xi$ by minimizing $\chi^2_\text{tot}$ --- the value of $\chi^2$ statistics between the model predictions for $Q_\text{II}(z)$ and $\tau_\text{es}$ and observations.  Because the `dark pixel' priors indicate only the lower bounds on the fraction of ionized hydrogen, we assume no contribution to $\chi^2_\text{tot}$ if the theoretically predicted $Q_{\rm HII}(z)$ are \emph{larger} than the mean values of `dark pixel' data at the corresponding redshift. 
Throughout this paper, we assumed the following values of the cosmological parameters consistent with the final \textit{Planck} results~\citep{Planck-2018}: $\Omega_\Lambda = 0.685$, $\Omega_\text{m} = 0.315$, $\Omega_\text{b} = 0.049$, $n = 0.961$, $\sigma_8 = 0.811$ and $h = 0.674$ (we also repeated our calculations using the \textit{Planck-XLVI} parameters  \cite{Planck-XLVI} and obtained that the our results do not  change significantly).

\section{Results}\label{sec:results}

\begin{figure*}
	\includegraphics[width=1.0\columnwidth]{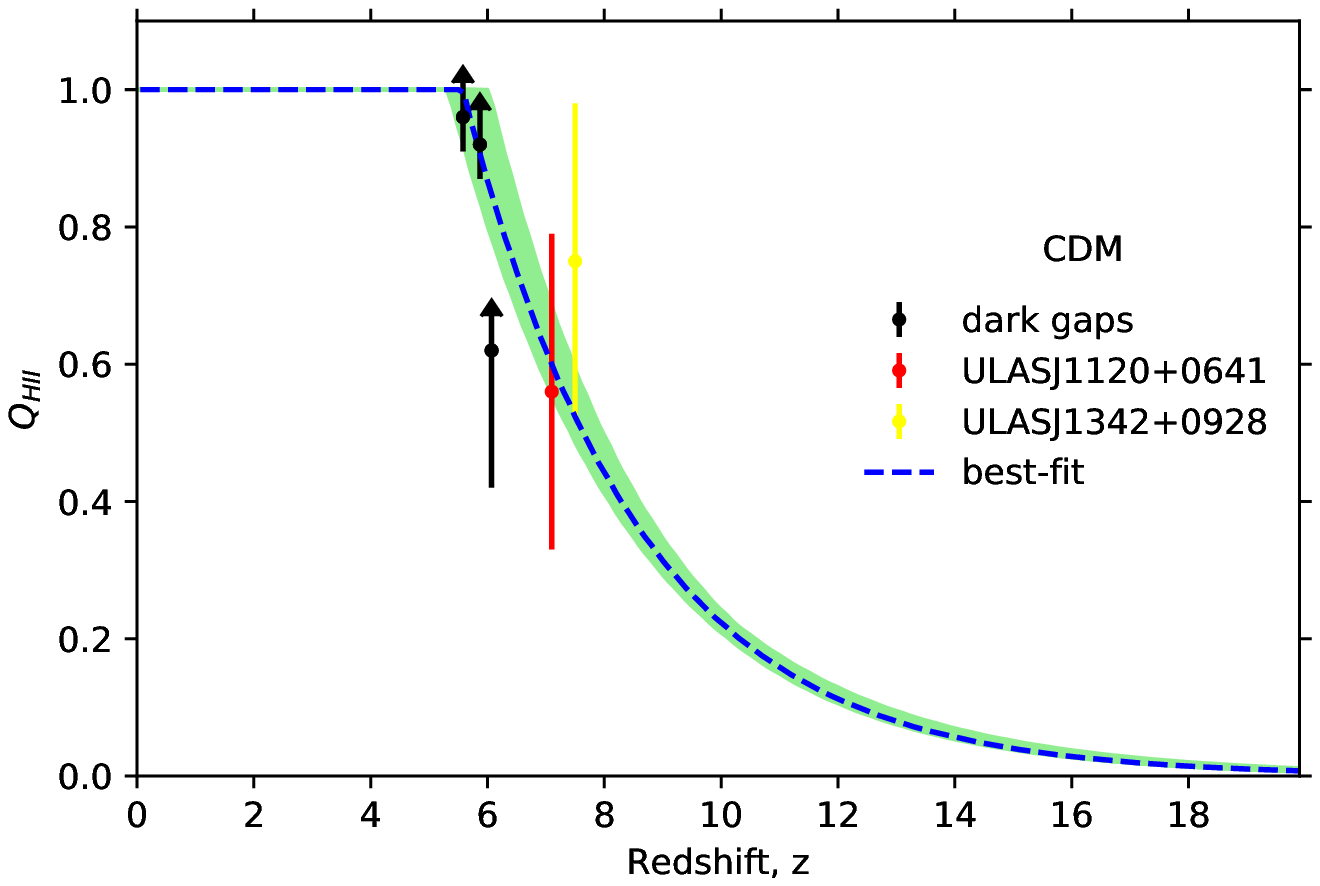}
	\includegraphics[width=1.0\columnwidth]{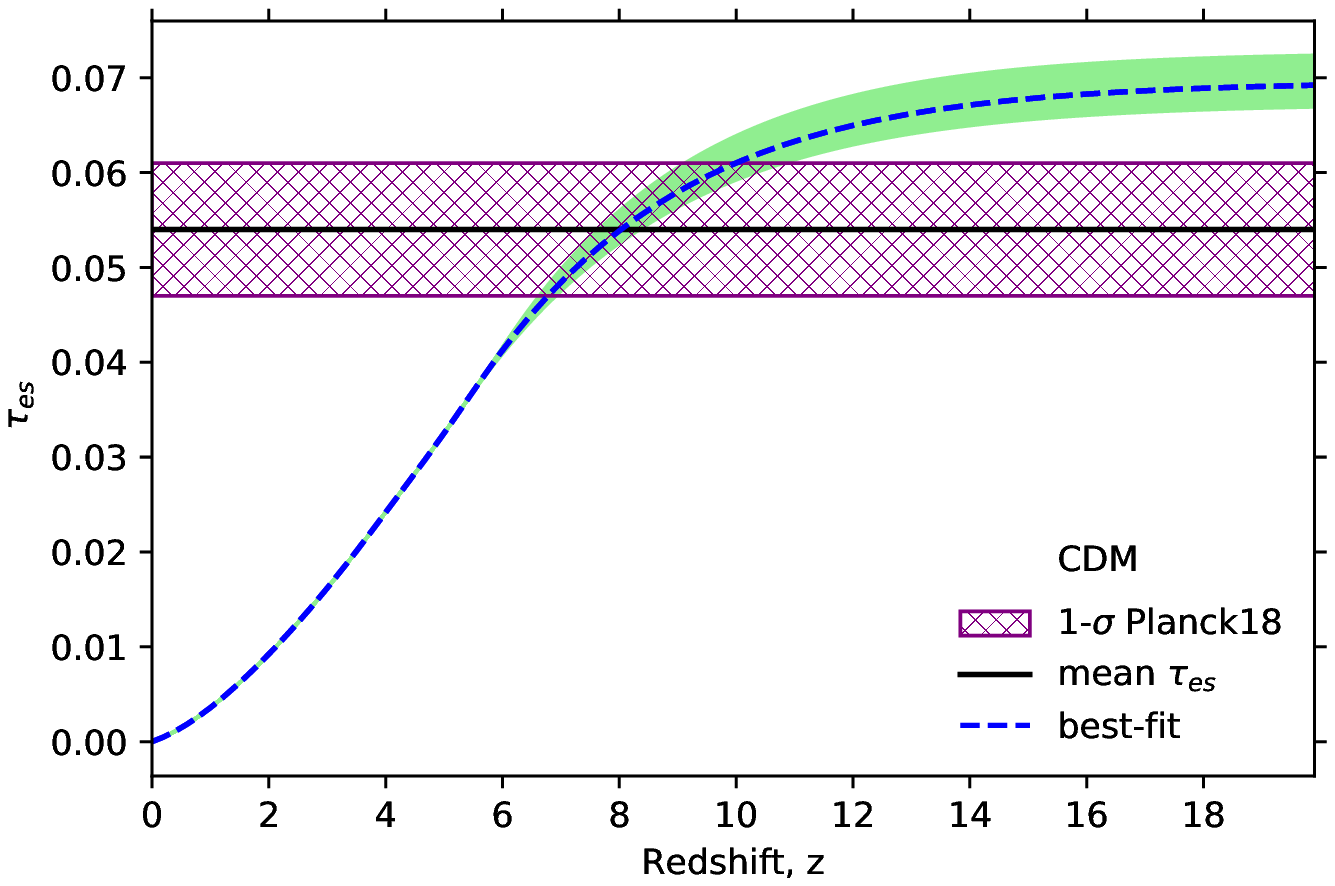}
	\includegraphics[width=1.0\columnwidth]{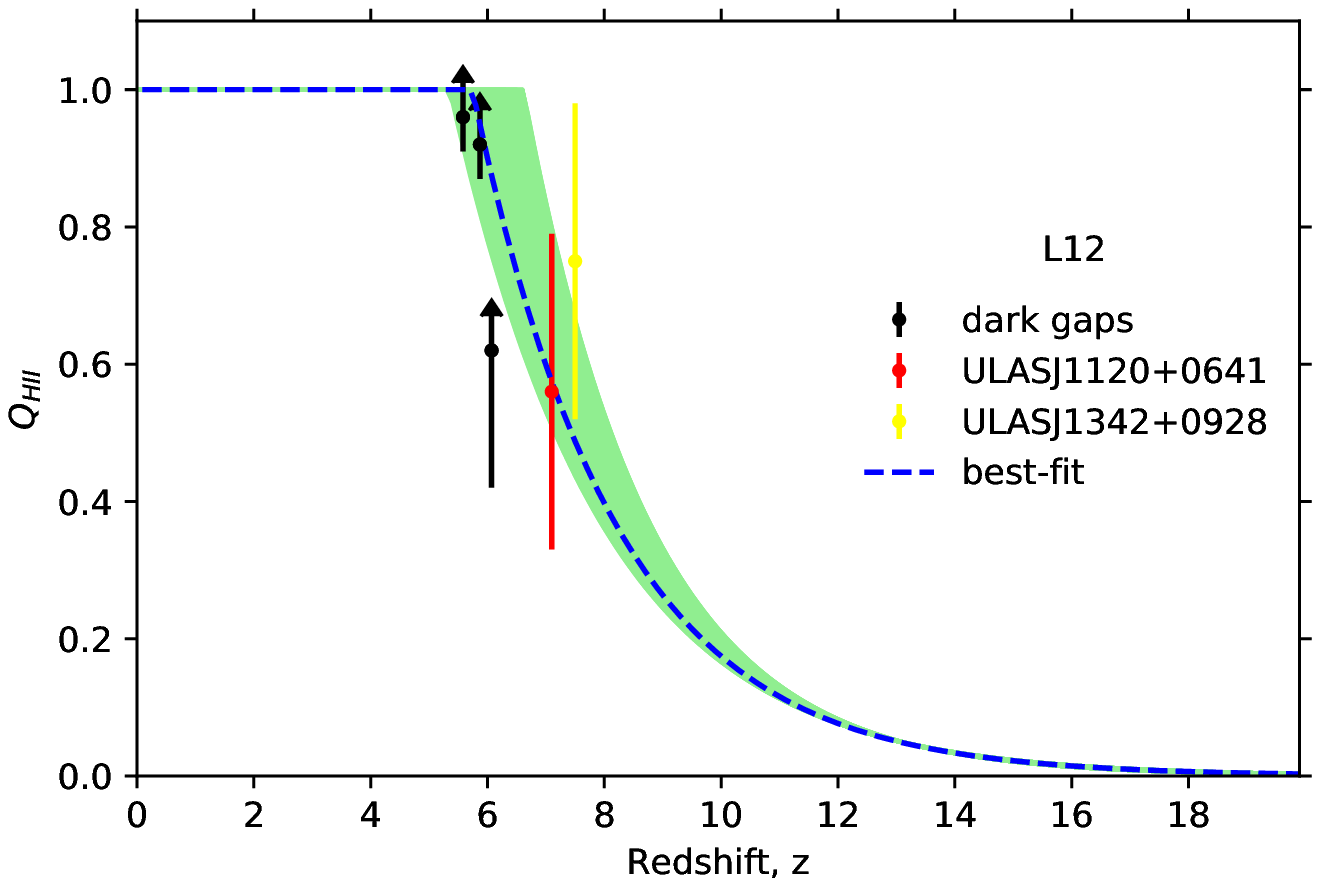}
	\includegraphics[width=1.0\columnwidth]{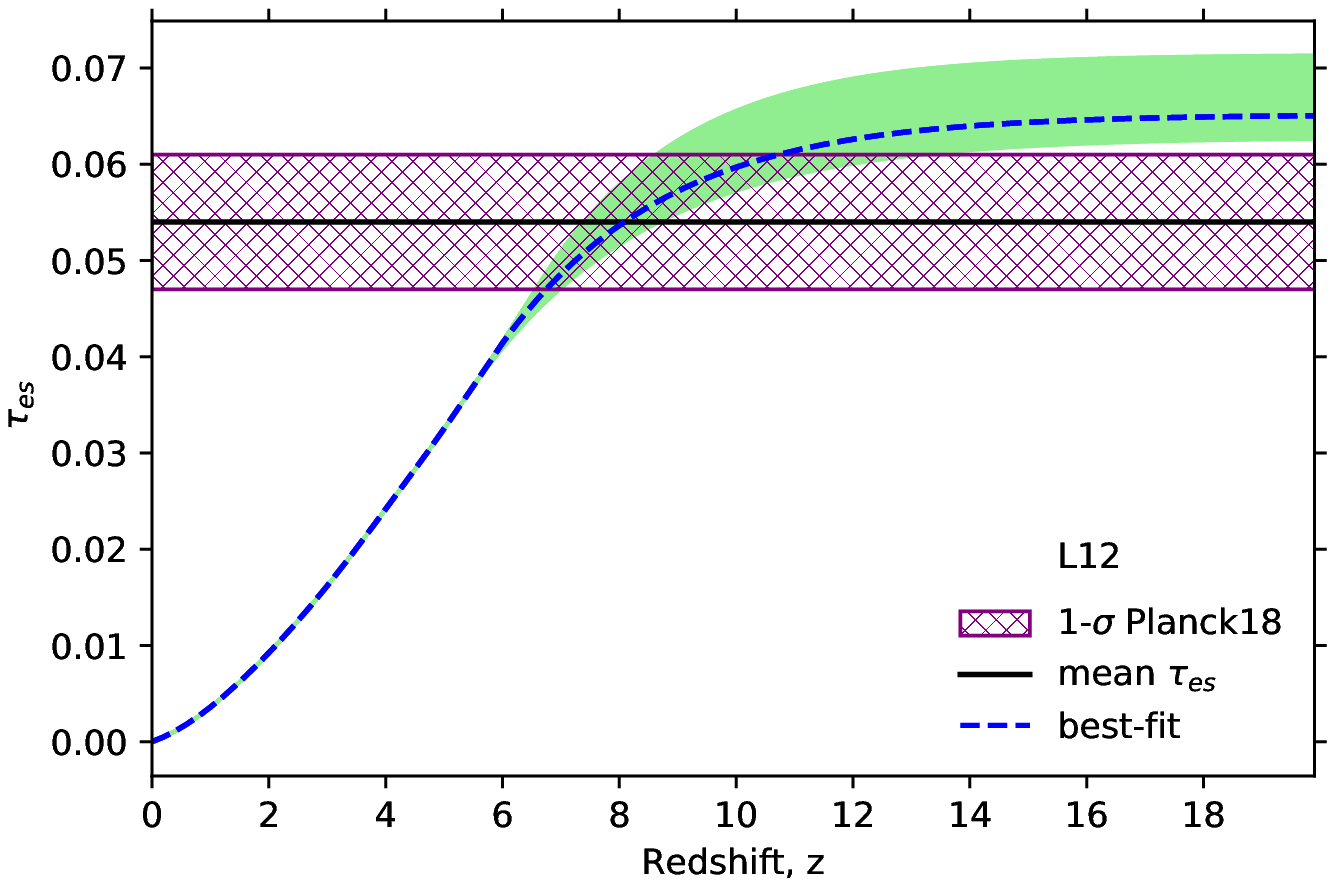}
	\includegraphics[width=1.0\columnwidth]{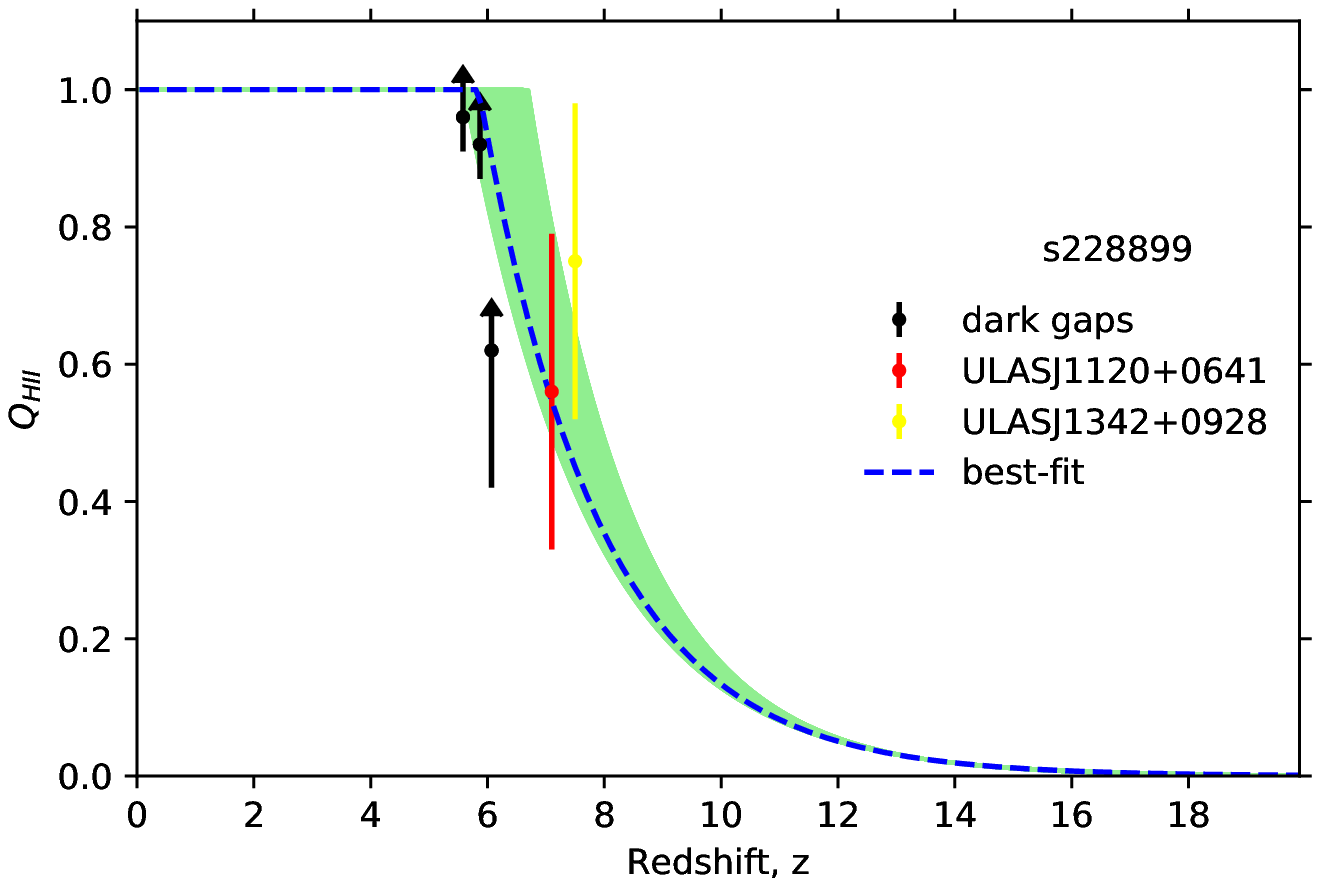}
	\includegraphics[width=1.0\columnwidth]{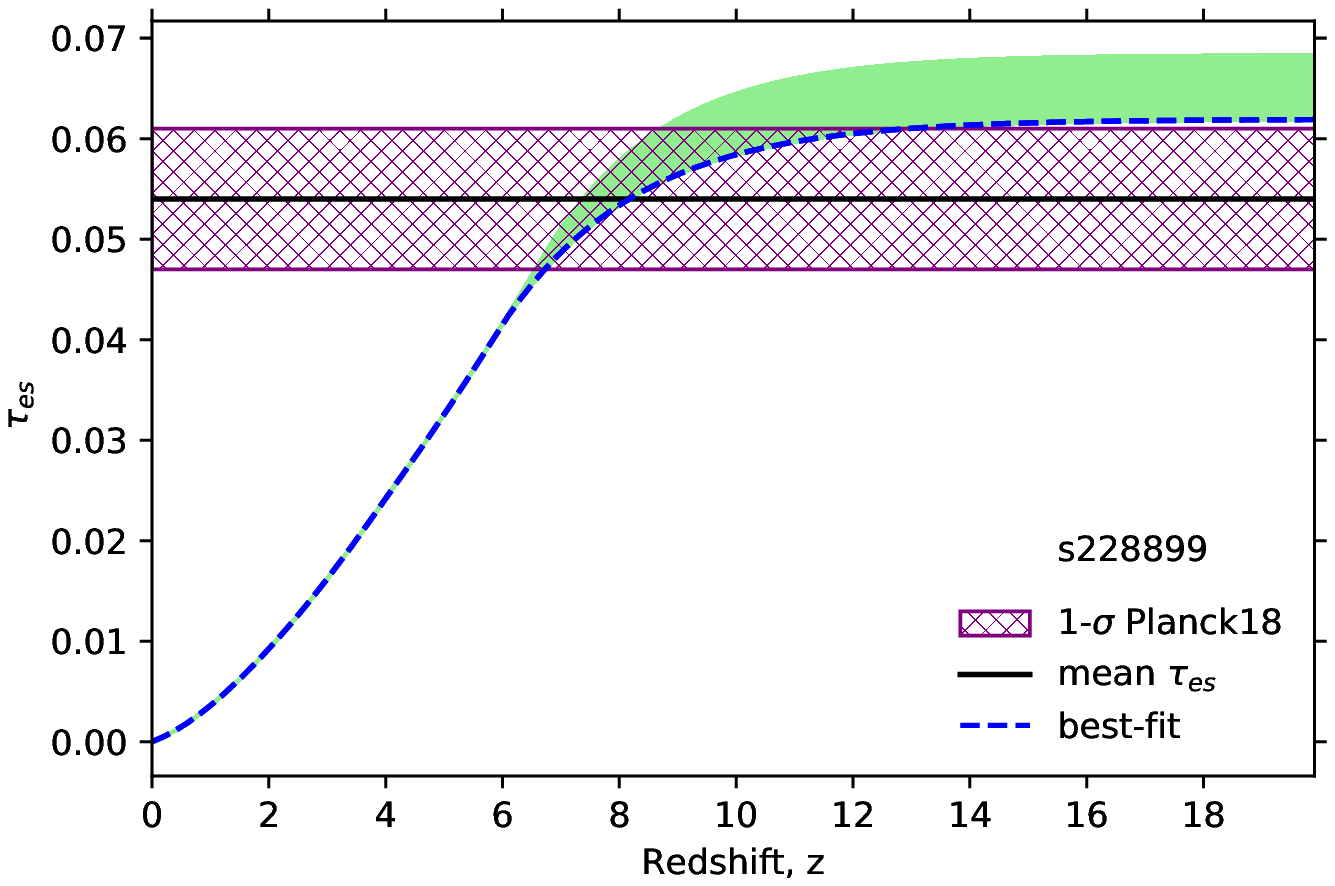}
   \caption[]{ \textit{Left}: Dependence of the ionized fraction $Q_\text{II}$  on redshift $z$  together with the priors from the analysis of dark pixels and damping wing absorption in quasar spectra. The upward arrows on priors from dark pixels mean that the only lower-bound errorbars can be obtained from this analysis.
   	
  \textit{Right}: Dependence of the optical scattering depth $\tau_\text{es}$ together with  measurements from \textit{Planck}.
  Blue dashed lines are the best-fit reionization histories for the indicated DM models. Light-green regions correspond to a range of reionization histories with $\Delta\chi_\text{tot}^2\leq4.61$ for each of our DM models. 
  }  
   \label{fig:results_T1e4}
\end{figure*}

We summarize the obtained results in Fig.~\ref{fig:results_T1e4}, which shows evolution of ionized volume filling fraction $Q_\text{II}(z)$ and the value of electron scattering optical depth $\tau_\text{es}$. The filled light-green regions correspond to the range of reionization histories with $\Delta \chi^2_{\text{tot}}=\chi^2_\text{tot}-\chi^2_{\text{tot, min}}\leq4.61$,\footnote{We fix the dark matter model during the calculation of the $\chi^2$ statistics. Therefore, we have 2 d.o.f. \@ for 2 varying free  model parameters ($\zeta$ and~$\xi$), and the $90\%$ confidence level corresponds to $\Delta \chi^2_{\text{tot}}=4.61$.} obtained for each of three dark matter models of our interest (CDM, L12 and s228899), together with the  observational priors described in detail in Sec.~\ref{sec:method}.  

We obtained statistically acceptable fit for all three models (given 4 degrees of freedom), resulting in $\chi^2_{\text{tot, min}} = 5.79$ for CDM, $\chi^2_{\text{tot, min}} = 3.80$ for L12, and $\chi^2_{\text{tot, min}} = 2.98$ for s228899 model. Although 7~keV sterile neutrino dark matter models appear to be more statistically preferred than the CDM model, the difference of $\chi^2_\text{min}$ is too small to make statistically robust preferences between these models.

\section{Discussion}\label{sec:discussion}

In this paper, the Pop II stars in galaxies are assumed as the main source of ionizing photons. In reality, other sources (such as active galactic nuclei (AGN) or possible  decays of dark matter particles) could produce ionizing photons during reionization epoch. However, the AGN contribution to reionization budget is usually thought to be sub-dominant, according to recent observational (see~\cite{Kashikawa:15,Ricci:16,Onoue:17,Parsa:17}) and numerical studies~\citep{Onorbe:17,Qin:17,Hassan:17}   (see, however, \cite{Giallongo:15, Madau:15, Chardin:16, Bosch-Ramon:18}). The contribution from the sterile neutrino DM decays is negligible since the expected lifetime of 7 keV sterile neutrino that may be responsible for the 3.5 keV emission line is at least two orders of magnitude greater than the  maximum lifetime of decaying dark matter particles that could re-ionize the universe near $z=6$~\citep{Liu:16,Oldengott:16,Slatyer:16,Poulin:16}.  \\

The recombination rate is proportional to the squared density of ionized gas; thus it is substantially boosted in high-dense neutral regions. The presence of neutral inhomogeneities should significantly affect the reionization history~\citep{Haiman:00,Iliev:04,Ciardi:05,McQuin:06, Finlator:12,Kaurov:13, Sobacchi:14, Park:16}. In this paper, we assumed that the  sinks of ionizing photons are the neutral minihaloes with masses in range from the Jeans mass up to the mass of the lightest galaxies and modelled them similarly to \cite{Rudakovskyi:16}.  For $\xi=3$$-$$9$ we obtain that,  on average,  two to three photons per hydrogen atom are required to ionize the Universe in the CDM cosmology, which is in good agreement with  \cite{Kaurov:13, So:13, Sobacchi:14}. In the warm dark matter models based on thermal relics or sterile neutrino, the formation of minihaloes is suppressed; therefore, the average number of recombinations  is reduced, see \citep{Yue:12, Rudakovskyi:16}. 

\section{Conclusions}\label{sec:conclusion}
By using the extended `Gold sample'~\citep{Greig:16b} of the existing measurements during the epoch of reionization, we found that both CDM and the 7~keV sterile neutrino models describe well the observational data on reionization.  The obtained difference between the $\chi^2$ statistics for CDM and sterile neutrino dark matter is $\Delta\chi^2_\text{tot} < 2-3$ depending of the sterile neutrino model of our choice\footnote{The difference of $\chi^2_\text{tot}$ values for L12 and s228899 sterile neutrino models is only 0.82, significantly smaller than the difference between CDM and each of sterile neutrino models (1.99 and 2.81 for L12 and s228899 models, respectively). Therefore, we expect that the fact that sterile neutrino models slightly better fit the available data on reionization to be independent of particular choice of sterile neutrino models. 
}. Taking into account the effects of baryonic feedback into reionization model could make this difference even smaller, in accordance with \cite{Dayal:17, Lopez-Honorez:17}.
Therefore, we conclude that existing observations of reionization do not allow to make any statistically significant distinction between the cold dark matter and 7~keV sterile neutrino dark matter models. Also, our results  qualitatively confirm the recent findings  \citep{Dayal:17, Lopez-Honorez:17} for the warm dark matter model with thermal relics.

\section*{Acknowledgements}

The authors thank the anonymous Referee, Oleg Ruchayskiy and Yuri Shtanov for their valuable comments and suggestions.
The work of A.R. was partially supported by the grant for young scientist's research laboratories of the National Academy of Sciences of Ukraine, the Program of Cosmic Research of the National Academy of Sciences of Ukraine, and grant 6F of the Department of Targeted Training of the Taras Shevchenko National University of Kiev under the National Academy of Sciences of Ukraine.
The work of D.I. is supported by a research grant from Carlsberg foundation.

\bibliographystyle{mnras}
\bibliography{preamble,reionization}

\bsp	
\label{lastpage}
\end{document}